**Dosimetric characterization of a nanophotonic scintillator and applications to real-time *in-vivo* total body irradiation dosimetry**

W. Jeffrey Zabel[1*], Dixin Chen[1], Louis Martin-Monier[2], Simo Pajovic[3,4], Shanhui Fan[5], Juejun Hu[2], Marin Soljačić[6], Lei Xing[1], Charles Roques-Carmes[5,7], M. Ramish Ashraf[1].

1. Department of Radiation Oncology, Stanford University School of Medicine, Palo Alto, CA, USA.
2. Department of Materials Science and Engineering, Massachusetts Institute of Technology, Cambridge, MA, USA.
3. Department of Mechanical Engineering, Massachusetts Institute of Technology, Cambridge, MA, USA.
4. Department of Applied Physics and Materials Science, California Institute of Technology, Pasadena, CA, USA.
5. E. L. Ginzton Laboratory, Stanford University, Stanford, CA, USA.
6. Department of Physics, Massachusetts Institute of Technology, Cambridge, MA, USA.
7. Institute of Science and Technology Austria (ISTA), Am Campus 1, 3400 Klosterneuburg, Austria.

*Corresponding author: wjzabel@stanford.edu

---

**Purpose:** Scintillators are widely used for both dosimetry and X-ray imaging. Recent advances in metasurface photonics and manufacturing have enabled a nanophotonic surface coating to be applied to conventional scintillators, which has been shown to significantly improve light yield. However, the dosimetric behavior of such coatings has not been established. We performed the first dosimetric characterization of a nanophotonic scintillator, quantitatively evaluating light enhancement, dose-rate dependence, linearity, and energy dependence relative to standard non-nanophotonic scintillators, and explored clinical application to real-time in-vivo total body irradiation (TBI) dosimetry as a potential initial use case.

**Methods and Materials:** A 4.5 cm × 1.5 cm cerium-doped yttrium aluminum garnet (YAG:Ce) scintillator—half patterned with the nanophotonic structure, half left unpatterned—enabled direct comparison between conventional and nanophotonic surfaces. The scintillator was placed in a 3D-printed light-tight box with an off-axis CMOS camera to measure light signal and irradiated with a clinical linear accelerator. Absolute dosimetry was done with dose-calibrated radiochromic film. For TBI, the scintillator was positioned at varying locations on an anthropomorphic phantom in a TBI booth, with signal measured by both CMOS and consumer-grade phone cameras under different room lighting conditions.

**Results:** The nanophotonic scintillator showed a 4.1× increase in signal and a 3.7× increase in contrast-to-noise ratio versus the conventional scintillator. Both scintillators exhibited dose-rate independence and linear dose response, with modest energy dependence for photon and electron beams. For TBI, the nanophotonic scintillator produced a clearly detectable signal with both CMOS and phone cameras, whereas the conventional scintillator signal was undetectable.

**Conclusions:** Nanophotonic structures significantly enhance light output of conventional scintillators without impacting their dosimetric properties. With further refinement, nanophotonic scintillators may enable real-time *in-vivo* TBI dosimetry using cost-effective, accessible camera systems. Future work should investigate this technology for improved dosimetry equipment and X-ray imaging detectors.

## Introduction

Scintillating materials, which emit light upon excitation by high-energy radiation, are integral to radiotherapy dosimetry, quality assurance (QA), and medical imaging [1]. Scintillators are most commonly used in CT and kV X-ray imaging detectors which convert X-rays into visible light, which is then detected by a CMOS photodiode [2]. They are also commonly used in electronic portal imaging devices (EPID) found on C-arm linear accelerators (linacs) which can be used for imaging guidance and high resolution dosimetric verification of the beam. For QA applications, scintillators are particularly useful in situations where conventional ion chambers exhibit undesirable properties such as in high-field MR-Linac environments or small-field dosimetry where volume averaging is a concern. Some innovative QA devices use scintillators for daily linac QA and isocenter verification [3] and brachytherapy [4], illustrating the increased use of scintillators for QA applications. Furthermore, scintillators have emerged as a premier candidate for ultra-high dose rate (FLASH) radiotherapy, as they circumvent the significant ion recombination challenges inherent to gas-filled detectors [5,6].

Despite the widespread integration of scintillators into clinical medical physics limited scintillator light yield necessitates the use of expensive, bulky light-amplification hardware, such as photomultiplier tubes, high-end photodiodes, and specialized cameras. These hardware requirements impose significant cost and spatial constraints on imaging and QA systems, ultimately hindering the widespread adoption of next-generation scintillation technologies [7]. Developing scintillators with enhanced light output could mitigate the need for costly detection hardware, thereby reducing the size and expense of QA and imaging devices. For kV imaging panels, such advancements could enable lower photon fluences to generate equivalent image quality, reducing patient dose. Alternatively, they could yield superior image quality at a standard dose. Advancements in the characteristics of scintillators to maximize light yield may thus drive significant improvements in both diagnostic imaging and dosimetry QA equipment [7].

Nanophotonics is a field of research that provides strategies to shape light by structuring matter on the scale of the optical wavelength or below [8]. Nanophotonics exploits geometry to engineer the local density of optical states, control the flow of light, enhance light-matter interaction, and control the angular, spectral, polarization, and temporal properties of optical fields. Canonical examples include photonic crystals [8], optical microcavities [9], nanoantennas [10], and metasurfaces [11, 12]. Over the past two decades, these concepts have enabled major advances in biosensing, integrated optics, nonlinear photonics, quantum optics, and compact imaging [12, 13, 14]. Some nanophotonic technologies have already crossed into commercial systems, from photonic-crystal-based optical components [15] to metasurface optics for wafer-scale sensing and polarization imaging [16]. However, similar technological advancements have not yet been realized in the field of medical physics, posing a significant opportunity for innovation and enhancement to our practice [7]. Roques-Carmes *et al* have recently proposed, and experimentally demonstrated, nanophotonic scintillators integrating nanophotonic structures such as subwavelength photonic crystal coatings atop bulk scintillator materials [17]. They demonstrated a 6-fold enhancement in the light output of a conventional scintillator by patterning of two-dimensional nanostructures on the surface [17]. Martin-Monier *et al* further developed a scalable and cost-effective manufacturing technique to create 4×4 cm nanophotonic scintillator surfaces improving the accessibility of this new technology [18]. Other nanophotonic designs have shown promises in enhancing timing, directionality, spatial resolution, and/or dose-efficiency of scintillators [19-24].

In this work, we performed the first dosimetric characterization of a cerium-doped yttrium aluminum garnet (YAG:Ce) nanophotonic scintillator. Although conventional YAG:Ce scintillators are widely used, this study was done to determine what benefits the nanophotonic surface coating provides as well as determine whether the coating will negatively affect the dosimetric properties of the scintillator. We also tested a potential first application of the scintillator in the clinic for total body irradiation (TBI) *in-vivo* dosimetry. TBI is a standard conditioning regimen for bone marrow transplantation in patients with leukemia or lymphoma, designed to kill residual cancer cells and provide necessary immunosuppression [25]. TBI is technically complex due to the large treatment distances, non-standard patient geometries, patient movement during treatment, and the critical need to maintain dose uniformity while sparing radiosensitive organs like the lungs. Due to these complexities, AAPM TG-29 specifies that *in-vivo* dosimetry should be performed to verify that the prescription matches what was delivered [26]. Currently, several technologies are utilized for this purpose, though each present with their own benefits and limitations. Thermoluminescent dosimeters (TLDs) and optically stimulated luminescent dosimeters (OSLDs) are the most common; however, both are passive "offline" detectors that require labor-intensive manual processing, preventing any real-time intervention if a dose discrepancy is detected during treatment [27,28]. Diodes and MOSFETs are also routinely used but are hampered by significant energy and angular dependencies, as well as sensitivity to temperature fluctuations and radiation damage, which necessitates frequent and complex recalibration [29,30]. We thus tested whether nanophotonic scintillators may be used as a surface dosimeter for real time TBI dosimetry. For TBI, long treatment distances result in a low dose rate, and thus weak light output, which makes it particularly challenging to use scintillators for *in-vivo* dosimetry. Work by Tendler *et al.* [31,32] and Niver *et al.* [33] evaluated scintillators for real time *in-vivo* TBI dosimetry however they used expensive intensified camera systems, with acquisitions gated to the linac pulses, to detect the weak light output. The increased light output of nanophotonic scintillators may enable accurate real time TBI dosimetry with more affordable and accessible camera systems.

## Methods

### Scintillator

A 4.5 × 1.5 × 0.05 cm semicircular YAG:Ce scintillator placed in a translucent container to prevent damage was used for all experiments (**Fig. 1A**). YAG:Ce is a commonly used inorganic scintillating material with maximum emission at 550 nm [34]. Half of the scintillator surface was left unpatterned, and the other half was patterned with a nanophotonic surface coating as described in Ref. [18]. This enabled a direct comparison of the output of the conventional vs. nanophotonic portions of the scintillator. The surface coating consists of a polymer coating with embedded chalcogenide glass and silica cladding (**Fig. 1BC**). These subwavelength structures were fabricated using an interference-lithography-defined master mold which enables patterning over centimeter-scale areas. The master texture was first transferred onto the scintillator by nanoimprint lithography: a polymer stamp replicated from the silicon master was pressed into a UV-curable polymer layer deposited on the scintillator and then exposed to UV to fix the pattern. A thin chalcogenide glass film was subsequently deposited by thermal evaporation. Finally, annealing above the glass transition temperature drove the film to reorganize into a highly ordered array of nanospheroids. The resulting photonic crystal had a subwavelength period of 450 nm. A thin $SiO_2$ capping layer was added to protect the chalcogenide nanostructures from further oxidation.

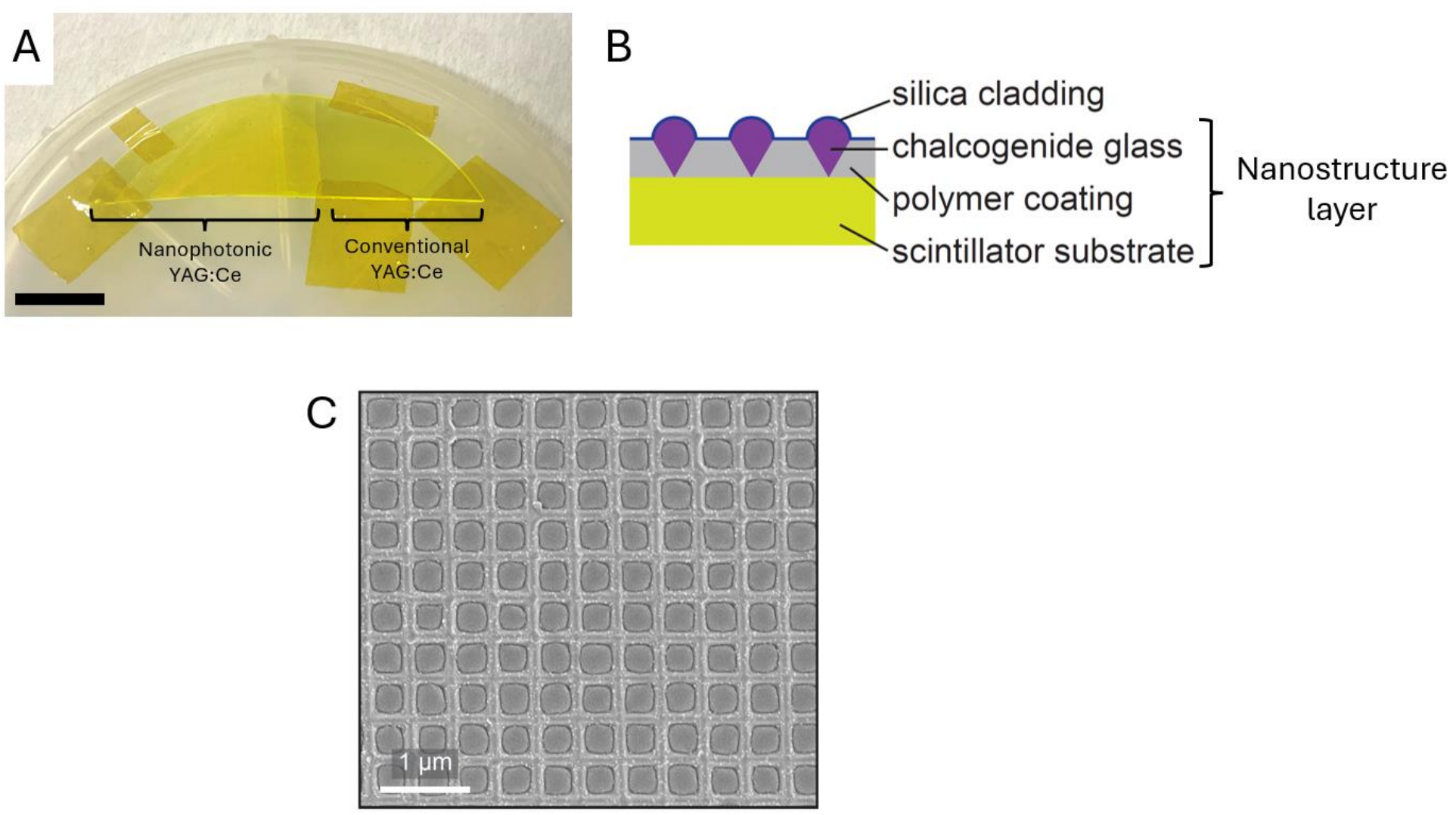

**Figure 1: Nanophotonic scintillator**. **A**, YAG:Ce scintillator with nanophotonic surface layer on half of the material in a translucent container to prevent damage. The scintillator is secured into a translucent container with yellow tape. **B**, A nanostructure layer was fabricated on top of the YAG:Ce scintillator (yellow) to enhance light output. **C**, scanning electron micrograph view of nanophotonic surface showing sub-wavelength size structures. **A**, scale bar = 1 cm.

Dosimetric characterization

To characterize the dosimetric properties of the scintillator, a 3D printed light tight box was used (**Fig. 2AB**). A monochrome CMOS camera (FLIR, Wilsonville OR, USA) with 25 mm fixed focal length lens (Edmund Optics, Barrington NJ, USA) was used to measure the scintillator light output (**Fig. 2CD**). The camera was positioned off-axis (outside of primary beam) and shielded with tungsten ball bearings to reduce radiation-induced speckle noise and damage. Unless otherwise stated, 200 ms exposure time and 5 frames per second were used.

A clinical linear accelerator (Varian Medical Systems Inc., Palo Alto CA, USA) was used to irradiate the scintillator. The scintillator was positioned at 100 cm source to surface distance and a $6 \times 11$ cm field size was used to fully encompass the scintillator (**Fig. 2B**). The linac was calibrated such that 1 monitor unit (MU) was equal to 1 cGy at the depth of maximal dose in water for each photon and electron energy. The gantry and collimator were set to 0° and varying energies were delivered with dose rate set to 600 MU/minute unless otherwise stated.

For absolute dose measurements, radiochromic film (Gafchromic EBT-XD, Ashland, Bridgewater NJ, USA) was first calibrated according to established protocols [35]. Films from the same batch were irradiated with known doses up to 50 Gy using 12 MeV electron beams under reference conditions (15×15 cm field size, 3 cm depth in solid water), with delivered dose corrected for daily output. Films were scanned 24 hours post-irradiation using a flatbed scanner (Epson 10000XL, Seiko Epson Corporation, Nagano, Japan) in 48-bit color mode at 150 dpi. Net optical density was calculated for the red channel and related to dose via a polynomial fit.

To convert delivered MU to absolute dose for each energy and modality (photons or electrons), 1,000 MU was delivered to a 1×3 inch radiochromic film placed over the scintillator. Using the edges of the films as fiducial markers, a rigid registration was used to map the film measured dose to the CMOS camera time series images of the scintillator. This provided a spatial dose distribution delivered to the scintillator (**Fig. 2E**) which could be related to the scintillator signal. This calibration protocol was performed for all photon and electron energies.

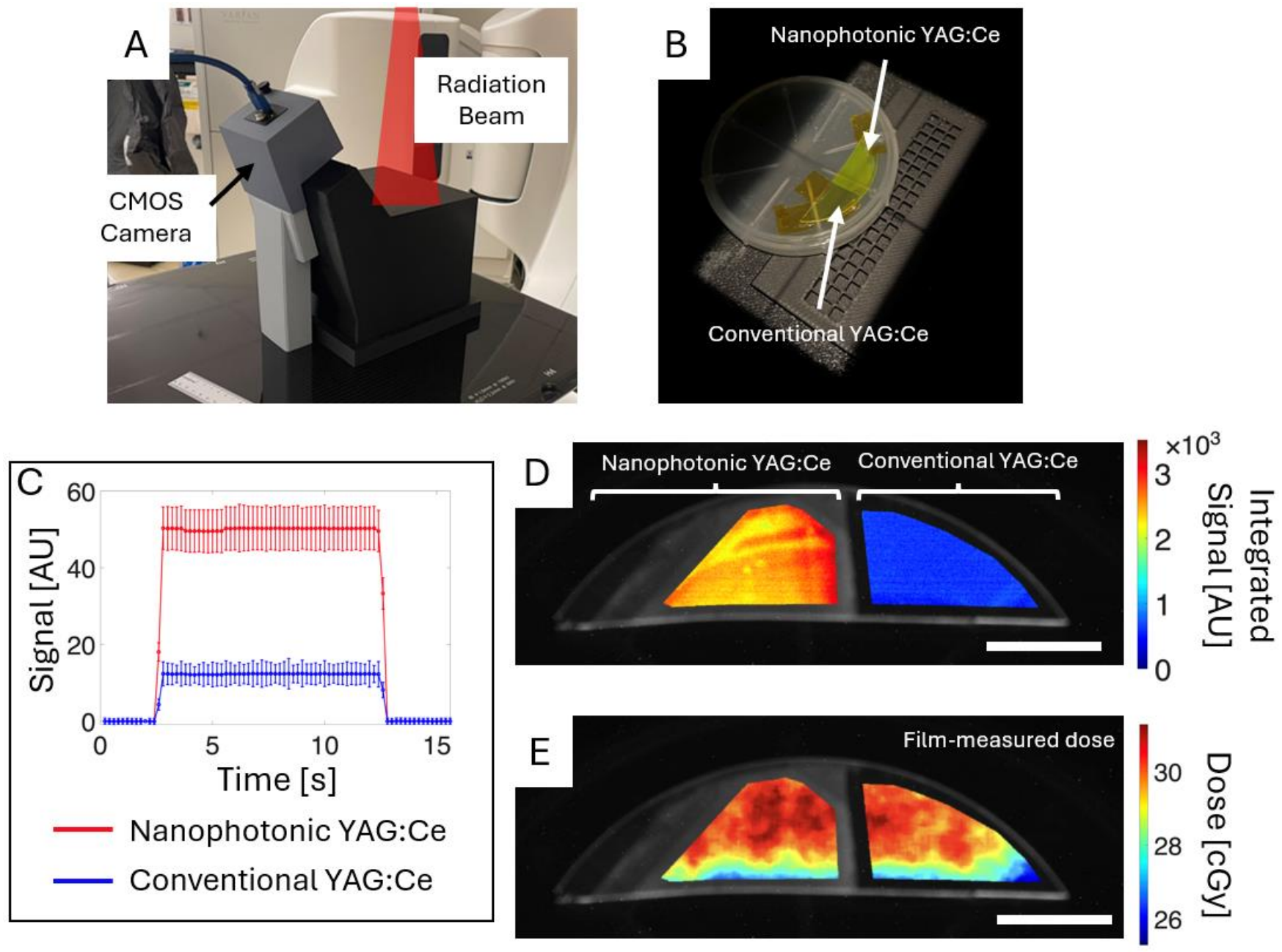

**Figure 1: Experimental setup for dosimetric characterization of scintillator.** **A**, A light-tight 3D printed enclosure with CMOS camera was used to measure scintillator signal during irradiation. **B**, The scintillator was positioned in the center of a $6 \times 11$ cm field. **C**, The light signal over beam delivery for conventional YAG:Ce (blue line) and nanophotonic YAG:Ce (red line). **D**, The integrated signal was measured for both conventional and nanophotonic YAG:Ce. **E**, The delivered dose was measured by dose-calibrated radiochromic film and coregistered to the scintillator images. **D**, **E**, Scale bar = 1 cm.

All analysis was performed in MATLAB R2025A (MathWorks, Natick MA, USA). For signal analysis, to remove X-ray speckle noise, a 2D median filter was applied to all time series images. Then the first 5 non-beam-on images were averaged and used as a background image which was subtracted from all time-series images. A region of interest (ROI) was manually applied to cover the conventional and nanophotonic scintillator portions to compare the signal output (**Fig. 2C**). Each pixel in the ROI was then integrated with respect to time (**Fig. 2D**) and the mean and standard deviation of the time-integrated spatial signal for each region were measured. The spatial absolute dose measurement from the radiochromic film was co-registered to the scintillator images and overlayed for correlation of the scintillator signal versus dose delivered (**Fig. 2E**).

The contrast to noise ratio was measured using equation (1) where $\overline{S_{scint}}$ is the average signal from the scintillator, $\overline{S_{bckg}}$ is the average signal in the background ROI with a similar area as the scintillator ROI, and $\sigma_{bckg}$ is the standard deviation of the background signal.

$$CNR = \frac{\overline{S_{scint}} - \overline{S_{bckg}}}{\sigma_{bckg}} \tag{1}$$

TBI Setup

An anthropomorphic phantom was placed in our clinic's TBI booth, and the scintillator was secured to three different positions: head, chest, and diaphragm. The CMOS camera was mounted on a tripod 140 cm from the phantom and 20 cm outside of the field edge to reduce direct camera irradiation. Exposure times of 500 ms were used with 2 frames per second. 3,000 MU 15 MV photons were delivered with the collimator at 45°, gantry 280°, 40×40 cm field size and 569 cm source to surface distance. A 1 cm plexiglass electron spoiler was used as is standard in our clinic. Room lights were at a level that was dim but still comfortable for a patient. The CMOS camera was then replaced with a consumer grade cellphone (iPhone 12 Pro Max, Apple, Cupertino CA, USA), and the room lights were turned off. 500 MU was delivered with the scintillator in the head position. The same data analysis pipeline that was used in the light tight box setup was used for the TBI analysis.

Statistics

For all plots, data points represent the mean and standard deviation of the time-integrated signal (or radiochromic film dose) from within the ROI of the scintillator unless otherwise stated. For the dose linearity and energy dependence experiment, the signal response as a function of dose was modelled using a zero-intercept linear equation of the form S = m·D. Fitting was performed for each energy using orthogonal distance regression to account for uncertainties in both the dose and signal measurements, with the slope uncertainty estimated using the effective variance method. The coefficient of determination, $R^2$, and the root mean square error, RMSE, were calculated for each fit. Energy dependence was assessed using an F-test comparing a single common slope fitted to all energies simultaneously against separately fitted slopes for each energy. A p-value greater than 0.05 indicates that allowing separate slopes does not significantly improve the fit, supporting the use of a single calibration slope across all beam energies.

## Results

### Light Output

6 MV, 100 MU, 600 MU/minute photons were delivered to the YAG:Ce scintillator. The integrated signal was measured and normalized to the conventional scintillator signal. The nanophotonic scintillator showed a 4.1-times enhancement in the light yield compared to the conventional scintillator.

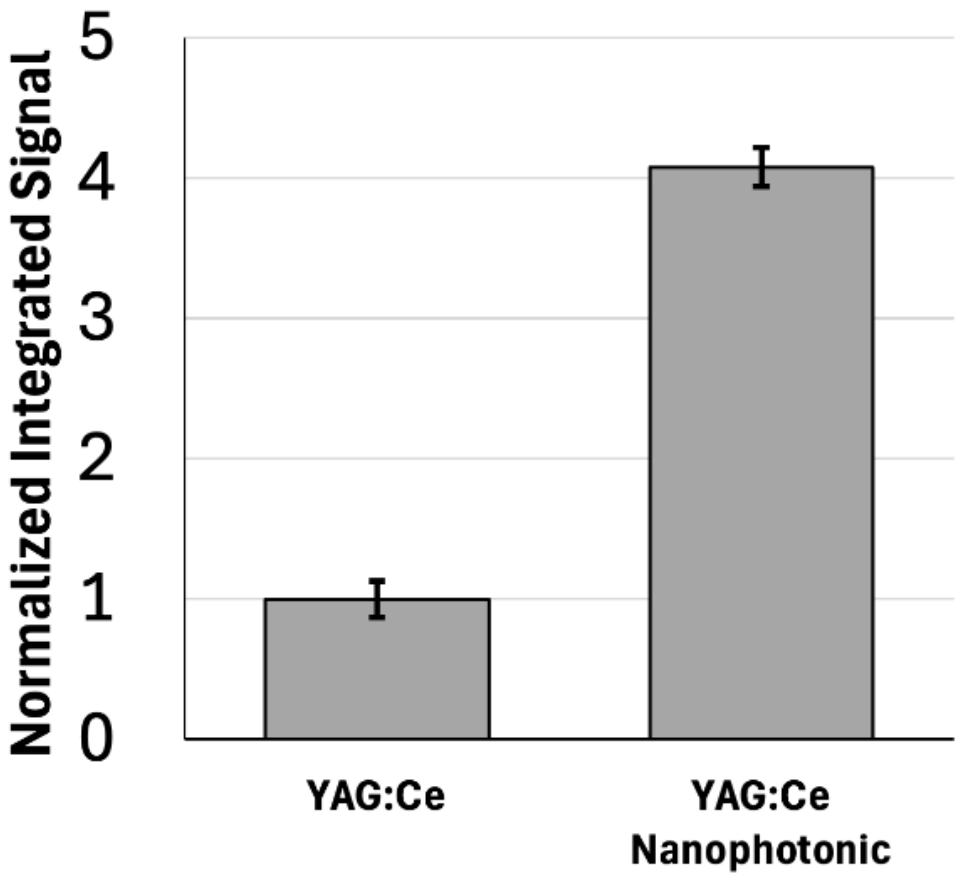

**Figure 3: Measurements of scintillator light output during irradiation.** Light output for the conventional and nanophotonic YAG:Ce scintillator for the same delivered dose. Measurements are normalized to the conventional YAG:Ce.

### Contrast-to-Noise Ratio (CNR)

CNR is an important metric to quantify how the light enhancement translates into improvements in camera acquisition parameters and resultant images. Specifically, if real time dosimetry is to be performed, a high frame rate (low camera exposure) with high CNR is required to maintain good temporal resolution and image quality. The CNR was thus measured for several different camera exposures during delivery of 6 MV, 100 MU, 600 MU/minute photons (**Fig. 4**). For both scintillators, the CNR improvements plateaued at 200 ms camera exposure. 200 ms camera exposure was thus used for all subsequent experiments since it offered the best CNR for the shortest camera exposure. The nanophotonic YAG:Ce exhibited a 3.7 time increase in CNR compared to the conventional YAG:Ce scintillator.

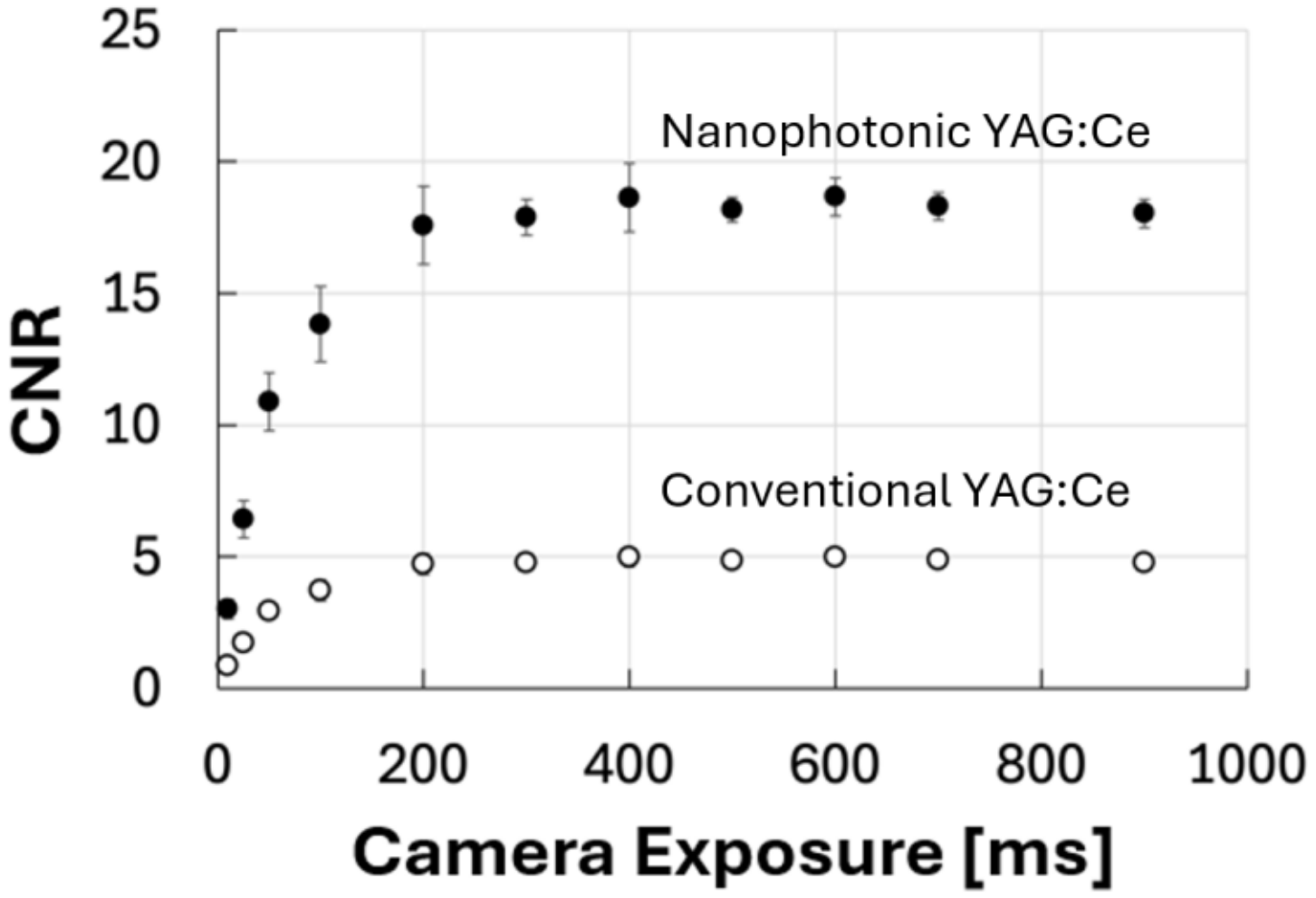

**Figure 4: Impact of scintillator enhancement on camera acquisition parameters and image quality.** The CNR was measured for conventional (hollow circles) and nanophotonic (solid circles) YAG:Ce for different camera exposures. Points represent the mean integrated signal in the ROI with standard deviation error bars (error bars for the conventional scintillator are too small to be seen).

Dose Rate Dependence

250 MU 10 MV flattening filter free (FFF) photons were delivered to the nanophotonic YAG:Ce scintillator at dose rates ranging from 400 MU/minute to 2,000 MU/minute. The measurements were normalized to the mean of all dose rate measurements and displayed in **Fig. 5**. Dose rate had no impact on scintillator light output with all measurements being within ±0.5% of the mean.

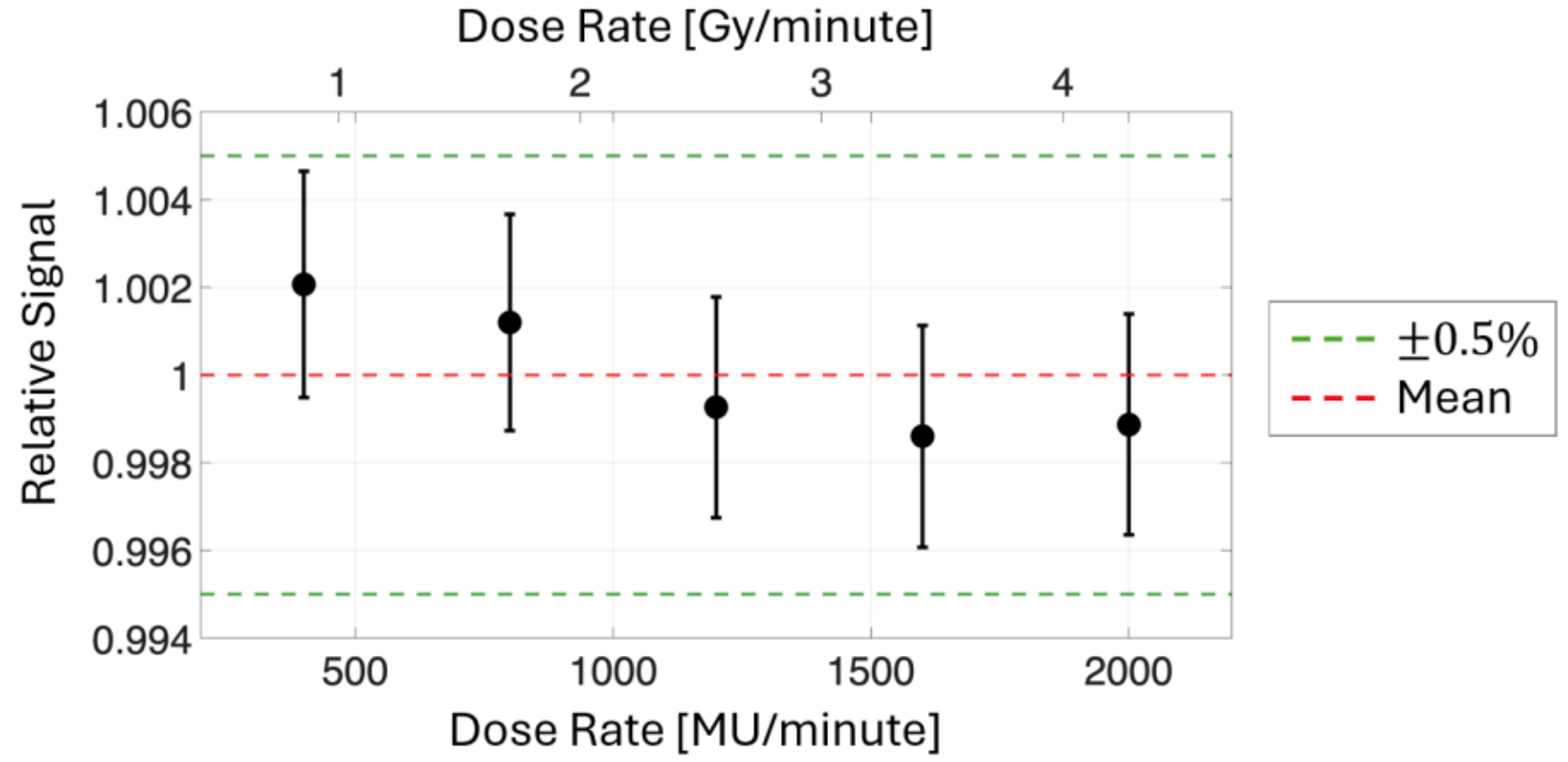

**Figure 5: Nanophotonic scintillator output for varying dose rates.** The scintillator outputs for different dose rates were normalized to their mean (red dotted line). Each point represents the mean ± standard error of the mean. All points were within ±0.5% of the mean (dotted green line).

Linearity and Energy Dependence

To assess dose linearity and energy dependence several doses were delivered to the nanophotonic scintillator with all photon (**Fig. 6A**) and electron (**Fig. 6B**) energies available at our clinic. The slope, $R^2$, and RMSE for all photon and electron energies are listed in **Table 1**. The nanophotonic scintillator exhibited a linear signal response to dose with $R^2 = 1.00$ for all photon and electron energies. The F-test yielded a significant result for photons ($F = 4758$, $p < 0.0001$) and electrons ($F = 600$, $p < 0.0001$), indicating that the signal response is significantly dependent on beam energy and separate calibration curves are required for each energy. For photons, all points were within 2.5% of the common fit, and for electrons all points were within 4% of the common fit. The non-nanophotonic YAG:Ce also exhibited a linear energy-dependent dose response however the slopes were approximately 4 times smaller than the nanophotonic scintillator due to the reduced light output (Appendix Fig. 1. and Appendix Table 1).

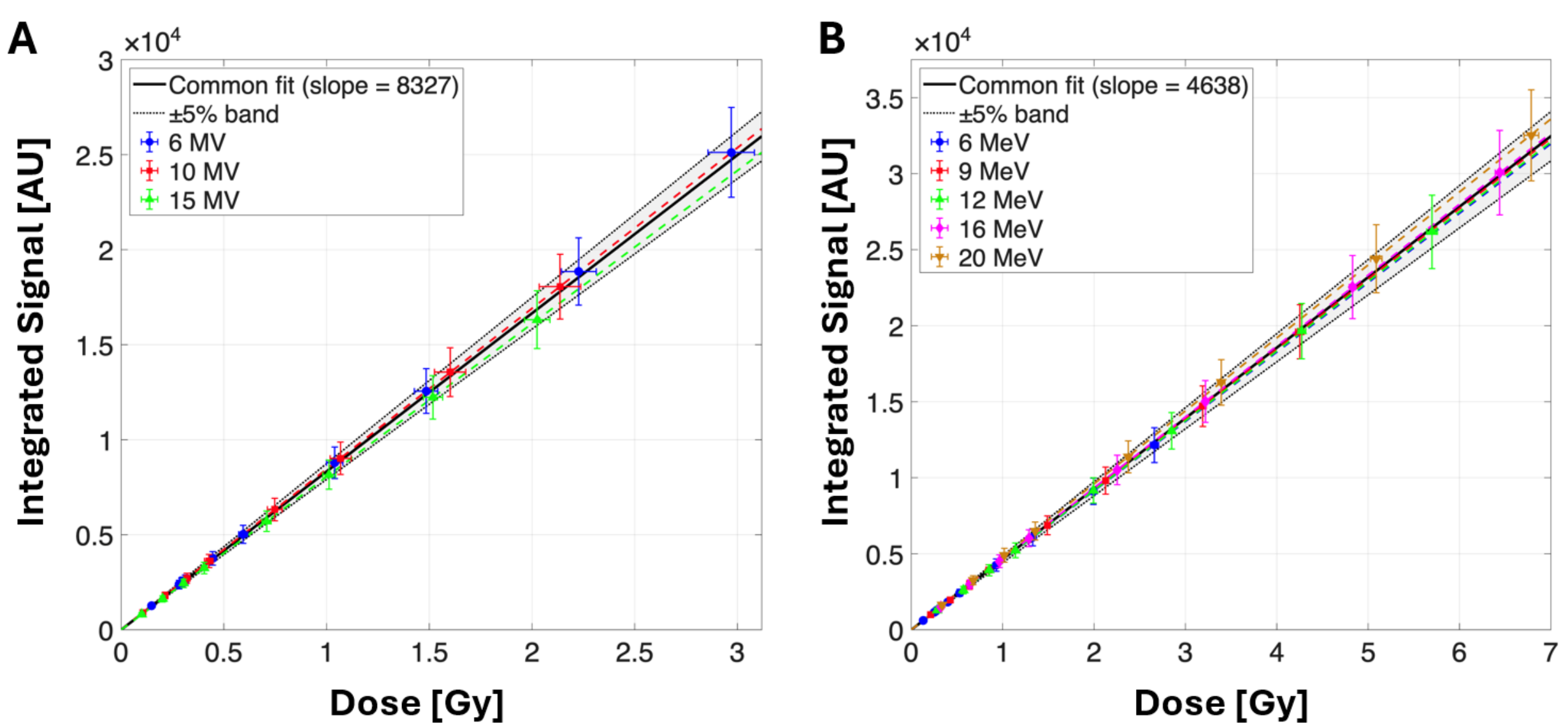

**Figure 6: Linearity and energy dependence of nanophotonic scintillator**. Various photon (**A**) and electron (**B**) energies were used to deliver a range of doses to the scintillator. For each energy a linear fit was applied as well as a linear fit for all energies combined to assess the energy dependence of the scintillator. A 5% band around the common fit was also plotted to help visualize the magnitude of the energy dependence.

| | Energy [MV or MeV] | Slope | $R^2$ | RMSE |
|---|---|---|---|---|
| **Photons** | 6 | 8454 ± 250 | 1.00 | 6.41 |
| | 10 | 8454 ± 320 | 1.00 | 5.63 |
| | 15 | 8053 ± 287 | 1.00 | 5.27 |
| | Common fit | 8327 ± 162 | 1.00 | 170.36 |
| **Electrons** | 6 | 4567 ± 129 | 1.00 | 3.56 |

| | 9 | 4615 ± 151 | 1.00 | 10.11 |
|---|---|---|---|---|
| | 12 | 4583 ± 151 | 1.00 | 29.21 |
| | 16 | 4665 ± 154 | 1.00 | 9.47 |
| | 20 | 4799 ± 158 | 1.00 | 19.64 |
| | Common fit | 4638 ± 66 | 1.00 | 237.58 |

**Table 1**: **Linear fit metrics for photon and electron signal vs. dose plots (nanophotonic scintillator).** The metrics (slope, $R^2$, and RMSE) for the linear fit (S = m·D) for all photon and electron energies as well as the common fits (all energies combined) are shown.

TBI Setup

An anthropomorphic phantom was set up in a TBI booth with CMOS camera placed outside of the booth and field (**Fig. 7A**). The integrated scintillator signal was measured during the TBI treatment with room lights dimmed but still comfortable for a patient. The scintillator signal was clearly detectable at all tested positions on the phantom: head, chest, diaphragm with CNR values of 2.29, 2.66, and 2.53 respectively (**Fig. 7B**). The CMOS camera was replaced with a conventional phone camera, and the integrated signal was measured with the room lights off (**Fig. 7C**). The scintillator was clearly detectable on the head of the phantom with CNR of 10.00. The higher CNR for the phone camera setup compared to the CMOS camera may be explained by the room lighting conditions in the two experiments (room lights on for CMOS camera measurements and room lights off for phone camera measurements). The conventional YAG:Ce signal was not distinguishable from background noise with either the CMOS or phone camera set ups.

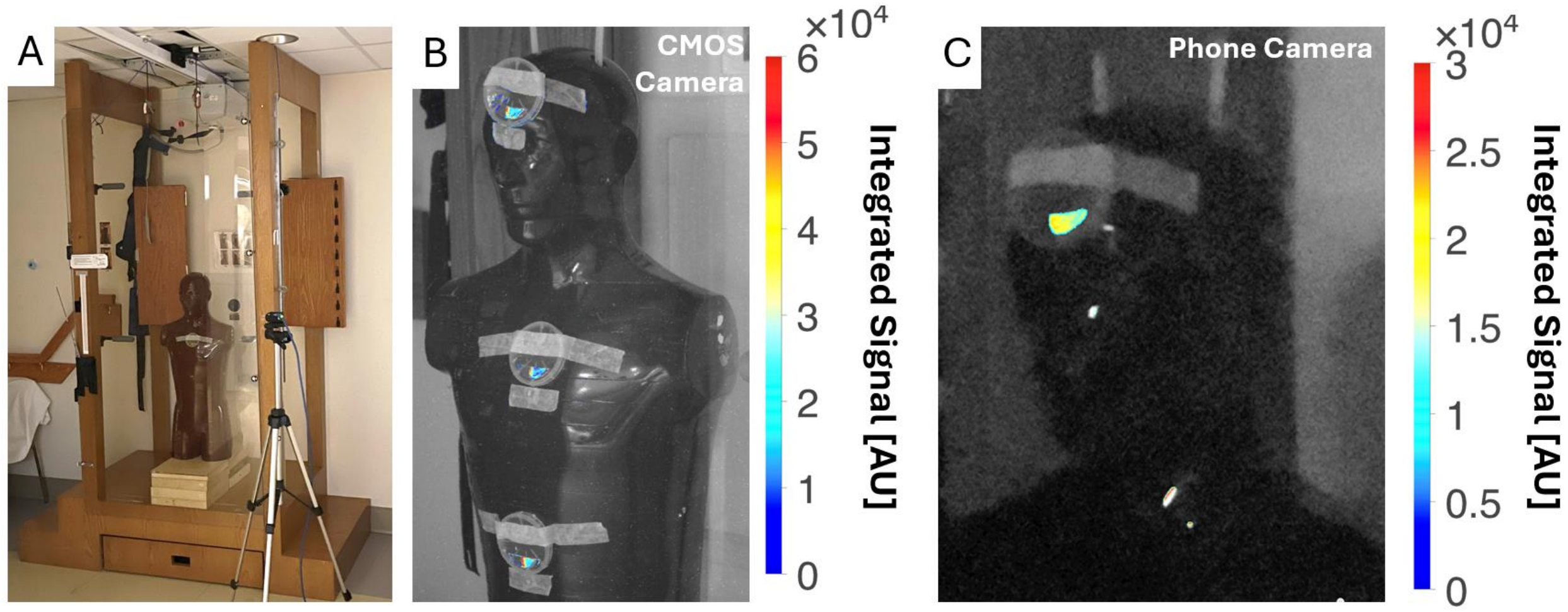

**Fig. 7: Nanophotonic scintillator for TBI dosimetry**. **A**, A anthropomorphic phantom was placed in a clinical TBI setup with CMOS camera positioned outside of the booth and field. **B**, Three different locations were tested (head, chest, diaphragm) and all positions showed a clearly detectable signal even with the room lights on. **C**, The CMOS camera was replaced with a consumer phone camera; with room lights off, the phone camera was clearly able to detect the scintillation signal.

## Discussion

Nanophotonic scintillators have the potential to transform radiation dosimetry and medical imaging due to their cost-effectiveness and significantly improved light output compared to conventional scintillating materials [18]. In this work, we performed the first clinical evaluation of this new technology by comparing it to conventional scintillators. We observed a 4.1-fold enhancement in light output of the nanophotonic scintillator compared to conventional scintillating materials (**Fig. 3**). This light enhancement is comparable to previous work [18]. A 3.7-fold increase in the CNR was also observed when comparing the conventional to nanophotonic scintillators down to 200 ms camera exposure time (**Fig. 4**). This indicates that high temporal resolution can be achieved while maintaining excellent signal enhancement. This is promising for real time dosimetry applications where high temporal resolution is necessary [36].

Although the nanophotonic scintillator exhibited significant light enhancement, we further evaluated whether the nanophotonic surface layer might negatively impact the dosimetric properties of the scintillator. We measured the integrated signal for several dose rates ranging from 400-2,000 MU/min and found a < 0.5% deviation in the signal response indicating that the scintillator was dose-rate independent (**Fig. 5**). This dose rate-independence is common for scintillators and typically translates to the FLASH dose-rate range [5]. FLASH dose rates were not evaluated here since more than enough signal can generally be obtained from conventional scintillators at these dose rates [5,6].

We found that the nanophotonic scintillator exhibited a linear dose response with modest energy dependence (**Fig. 6**). This can be explained as an intrinsic property of the chosen scintillator (YAG:Ce) and not the nanophotonic surface layer since the conventional scintillator also exhibited modest energy dependence (Appendix Fig.1 and Table 1). This energy dependence was also noted in previous works in the kV energy range [18]. It was found that all points fell within 2.5% of the common fit for photons, and within 4% of the common fit for electrons. Due to the overall minor energy dependence, this may be ignored depending on the required accuracy for the dose measurement. Nevertheless, it is recommended that an energy-specific calibration curve be used for future work using this scintillator.

The slope of the signal vs. dose line was larger for photons than electrons (8327 ± 162 for photons and 4638 ± 66 for electrons) (**Fig. 6**). This is perhaps counterintuitive since electrons may be expected to produce a stronger scintillation signal for the same dose as they are directly ionizing whereas photons must produce secondary charged particles to deposit their dose. To explain this, we must first consider the geometry of the setup for measuring the absorbed dose – radiochromic film was placed over the scintillator and 1,000 MU of either photons or electrons was delivered. A calibration factor to convert MU to absolute dose was then calculated for each photon and electron energy. In a megavoltage photon beam, dose is deposited by secondary electrons liberated through photon interactions, and at a point in air without overlying buildup material the local secondary electron fluence is small and strongly dependent on the detector itself. The radiochromic film is thin and approximately water-equivalent, so it generates few secondary electrons internally and reports a correspondingly low dose per MU. The 0.5 mm YAG:Ce layer, by contrast, has a much higher effective atomic number and density, producing substantially more secondary electrons within its own volume and thereby self-establishing a partial buildup that enhances energy deposition relative to the film. The photon-derived dose calibration therefore underestimates the energy absorbed by the scintillator, inflating the apparent signal-per-dose slope. For the megavoltage electron beam, both detectors are exposed to a directly incident charged particle fluence and neither relies on internal secondary electron generation, so the film and scintillator sample comparable dose conditions and the resulting slope more faithfully reflects the intrinsic scintillator response. Although this setup is useful for comparing different photon energies with each other, and different electron energies with each other, it is not useful to compare the electron and photon slopes. To perform a direct comparison between the signal produced by electrons and photons, the scintillator could be placed at reference conditions inside of a water tank however this would lead to a reduced ability to measure the scintillator signal due to scatter and attenuation in the water. Alternatively, a fiber optic cable may be used, however this would require a complete redesign of the detector.

We further performed measurements of the detector in a mock TBI setup with an anthropomorphic phantom (**Fig. 7**). The purpose of this was to quantify the light enhancement and detectability of the scintillator using different cameras and room lighting conditions. The scintillator was clearly detectable with a CMOS camera and room lights on (**Fig. 7B**). Even a consumer-grade phone camera could clearly detect the signal although this measurement was performed with the room lights off which would not be ideal for patient comfort (**Fig. 7C**). In both experiments, the cameras were placed outside of the plastic spoiler which could reduce the signal. With further optimizations to this setup, we plan to develop a real-time *in-vivo* dosimetry system for TBI applications. Future work will focus on camera position optimization to achieve the best signal while maintaining view of the scintillator on the patient's body and measurements of absolute dose calibration factors for different positions of the scintillator. Even simple modifications such as the application of a reflective coating on the back of the detector have the potential to double the scintillator light output. To further characterize the response of the scintillator we will measure the impact of radiation damage as well as its temperature and angular dependencies. The cost-effective scintillator production method [18] and camera systems would make this system a potentially affordable and accessible option for clinics compared to other proposed systems using expensive intensified cameras [31-33]. The proposed system could eliminate many of the issues with current TBI detector technologies such as OSLD's or TLD's which have a long readout time delay [27,28] and MOSFET's or diodes which have significant energy, temperature, and angular dependencies [29,30].

One potential drawback of the current nanophotonic scintillator is that it is not tissue equivalent. Although this is not expected to cause an issue for applications to TBI *in-vivo* dosimetry (thin scintillator is not expected to perturb the beam significantly) it may pose a problem if this technology is applied for small field dosimetry. Plastic scintillators are increasingly being used for small field dosimetry applications due to their small size, high resolution, and tissue equivalence. The nanophotonic scintillator used in this study is based on the YAG:Ce scintillator which may lead to inaccurate dosimetry at small field sizes due to perturbation of the local electron fluence. A similar nanophotonic strategy could in principle be applied to plastic scintillators, either by imprinting a nanoscale texture directly into the polymer surface or by depositing a patterned high-index coating on top of the plastic substrate. The expected enhancement may be more modest than in YAG:Ce, since common plastic scintillators have lower refractive indices around ($n \sim 1.5$-$1.6$), reducing the available index contrast for photonic structuring. Practical challenges include the lower softening temperature of plastics, compatibility with solvents and vacuum deposition, and preserving the scintillator's optical transparency during processing. For large-area plastic scintillators, a recent alternative is to use controlled random surface texturing rather than periodic nanophotonic patterning, which can improve light extraction through a scalable and low-cost process [37].

Outside of small-field and TBI dosimetry, we envision this technology being applied to X-ray medical imaging. Current X-ray imaging systems such as CT machines rely on scintillators to convert X-rays that have passed through the patient into light which is then detected by a photodiode. Nanophotonic scintillators, which exhibit a 4-fold light enhancement, may enable either (1) reduced dose to the patient since fewer X-rays would be required to produce the same quality of image, or (2) significantly improved image quality for the same radiation dose. Future research in this field could lead to significant improvements to our radiation dosimetry and imaging technologies.

## Conclusion

Recent advances in nanophotonics and scalable nanofabrication capabilities have provided unique opportunities to produce cost-effective cm-scale nanophotonic scintillators that produce significant light yield enhancements. We performed the first dosimetric characterization of this scintillator and showcased its potential for clinical applications in dosimetry for radiation oncology. By comparing the conventional and nanophotonic scintillators with each other, we found that, the nanophotonic scintillator had improved light output without impacting the dosimetric properties such as dose-rate, linearity, and energy dependence. We further evaluated its applications for real time TBI *in-vivo* dosimetry and found that the light signal can be easily detected using cost effective camera systems and practical room lighting and set up conditions. Future work will focus on further characterization of the scintillators dosimetric properties, refinement of the TBI *in-vivo* dosimetry system, and potential applications of this technology to improve X-ray imaging panels and QA equipment.

## Acknowledgments

This work was supported in part by the U.S. Army Research Office through the Institute for Soldier Nanotechnologies under Cooperative Agreement Number W911NF-23-2-0121, in part by the Stanford University Bio-X seed grant, and in part by the National Institutes of Health (grant

#1R01CA223667). SP was supported by the NSF Graduate Research Fellowship Program (GRFP) under grant no. 2141064.

**Data availability**

The data that support the findings of this study are available from the corresponding author upon reasonable request.

## Appendix

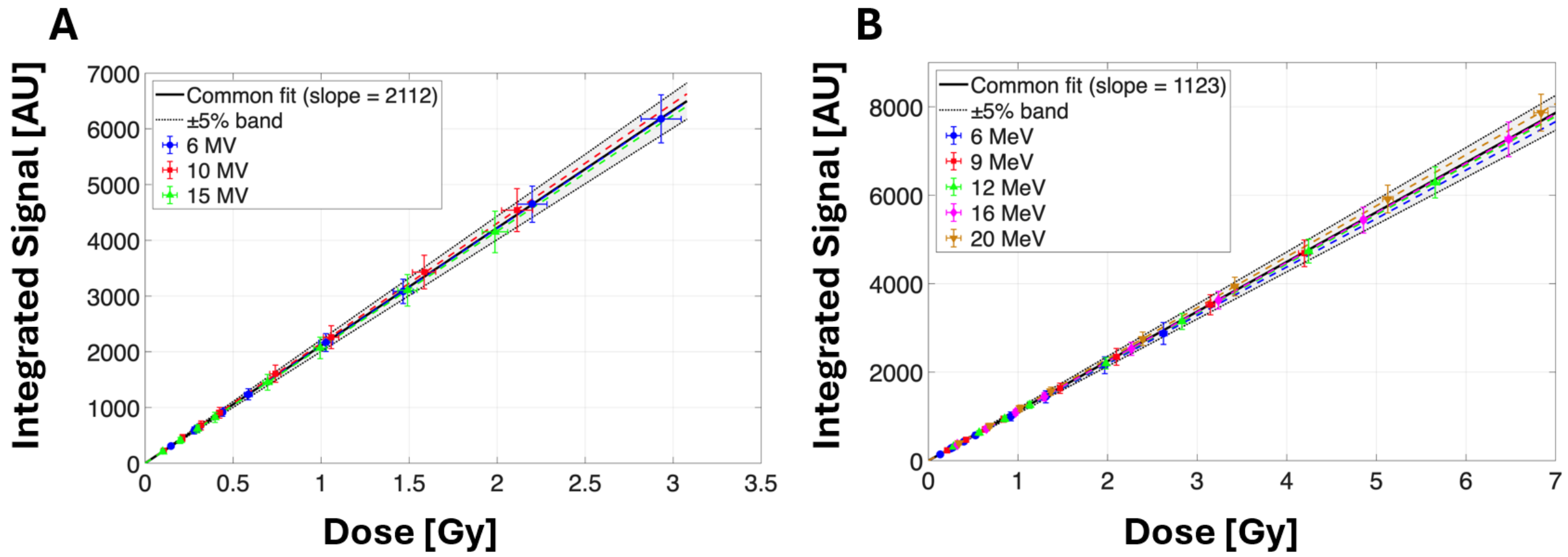

**Appendix Figure 1: Linearity and energy dependence of conventional scintillator**. Various photon (**A**) and electron (**B**) energies were used to deliver a range of doses to the scintillator. For each energy a linear fit was applied as well as a linear fit for all energies combined to assess the energy dependence of the scintillator. A 5% band around the common fit was also plotted to help visualize the magnitude of the energy dependence. For photons, the slopes are significantly different, and the response is energy dependent ($F = 150$, $p < 0.05$). For electrons, the slopes are significantly different, and the response is energy dependent ($F = 375$, $p < 0.05$).

| | **Energy [MV or MeV]** | **Slope** | **$R^2$** | **RMSE** |
|---|---|---|---|---|
| **Photons** | 6 | 2106 ± 57 | 1.00 | 4.24 |
| | 10 | 2153 ± 86 | 1.00 | 7.60 |
| | 15 | 2081 ± 83 | 1.00 | 5.27 |
| | Common fit | 2111 ± 41 | 1.00 | 23.27 |

| **Electrons** | 6 | 1095 ± 37 | 1.00 | 2.91 |
|---|---|---|---|---|
| | 9 | 1119 ± 32 | 1.00 | 2.55 |
| | 12 | 1112 ± 26 | 1.00 | 6.78 |
| | 16 | 1120 ± 25 | 1.00 | 6.13 |
| | 20 | 1152 ± 25 | 1.00 | 5.27 |
| | Common fit | 1123 ± 12 | 1.00 | 44.51 |

**Appendix Table 1**: **Linear fit metrics for photon and electron signal vs. dose plots (conventional scintillator).** The metrics (slope, $R^2$, and RMSE) for the linear fit ($S = m \cdot D$) for all photon and electron energies as well as the common fits (all energies combined) are shown.